\def\d#1/d#2{ {\partial #1\over\partial #2} }


\def\pdr{\partial}

\def\eps{\epsilon}


\newcount\eqnumber
\def\beq{ \global\advance\eqnumber by 1 $$ }
\def\eeq{ \eqno(\the\eqnumber)$$ }
\def\n{\global\advance \eqnumber by 1\eqno(\the\eqnumber)}
\def\puteqno{
\global\advance \eqnumber by 1 (\the\eqnumber)}


\def\ifundefined#1{\expandafter\ifx\csname
#1\endcsname\relax}
 \newcount\sectnumber \sectnumber=0
\def\sect#1{ \advance \sectnumber by 1 {\it \the \sectnumber. #1} }

\newcount\refno \refno=0  
\def\[#1]{
\ifundefined{#1}
\advance\refno by 1
\expandafter\edef\csname #1\endcsname{\the\refno}\fi[\csname
#1\endcsname]}
\def\refis#1{\noindent\csname #1\endcsname. }

\def\label#1{
\ifundefined{#1}
\expandafter\edef\csname #1\endcsname{\the\eqnumber}
\else\message{label #1 already in use}
\fi{}}
\def\(#1){(\csname #1\endcsname)}
\def\eqn#1{(\csname #1\endcsname)}

\baselineskip=13pt
\magnification=1000
\hoffset = 1in
\voffset = 1.25in
\hsize = 4.7in
\vsize = 7in
\def\BEGINIGNORE#1ENDIGNORE{}


\hfuzz = 0.5in
\def\label#1{
\ifundefined{#1}
\expandafter\edef\csname #1\endcsname{\the\eqnumber}
\else\message{label #1 already in use}
\fi{}}
\def\(#1){(\csname #1\endcsname)}
\def\eqn#1{(\csname #1\endcsname)}

\centerline{\bf RENORMALIZED PATH INTEGRAL IN QUANTUM MECHANICS}
\vskip 1.5pc 
\centerline{R.J. HENDERSON and S. G. RAJEEV}

\centerline{Henderson or Rajeev@urhep.pas.rochester.edu} 
 \centerline{\it Department of Physics and Astronomy, University of
Rochester, Rochester, NY 14627}
 
\vskip 1pc 
                          
{\narrower \baselineskip = 10pt \sevenrm We obtain direct, 
finite, descriptions of a renormalized quantum mechanical
system with no reference to ultraviolet cutoffs and running coupling constants,
in both the Hamiltonian and path integral pictures. The path integral
description requires a modification to the Wiener measure on continuous paths
that describes an unusual diffusion process wherein colliding particles
occasionally stick together for a random interval of time before going their
separate ways.\par} 

\vskip 1.5pc 
\noindent{\bf 1 \hskip 0.2pc Introduction} 
\vskip 1pc 
\noindent The presence of ultraviolet divergences in the quantum field theories
of the standard model, and the need for an awkward renormalization procedure
to make these theories well-defined, might be viewed as evidence that quantum
field theory is not the proper framework for a fundamental theory of elementary
particles. Some exotic and finite theory, such as string theory, might be more
conceptually accurate and less mathematically cumbersome. On the other hand, a
more conservative point of view is that renormalizable interactions can be
given a finite description, which avoids renormalization, without the necessity
of discarding the framework of quantum field theory. 

It would be aesthetically pleasing and probably computationally useful to
construct a theory of renormalizable interactions which is finite at the outset
and does not require the seemingly artificial limiting procedures of
renormalization to be well-defined. Such a description is presently beyond our
grasp, but here we construct a finite description of a renormalizable quantum
mechanical system that suggests that quantum field theory can possibly 
accomodate renormalizable interactions through a choice of the Hamiltonian 
domain rather than the
addition of an interaction term in the Lagrangian. In the case we examine, we
thus find evidence that we can eliminate the need for renormalization
altogether by taking the conservative point of view which requires no exotic
replacement for quantum field theory. An analagous
{\it finite} description of renormalizable interactions in quantum field theory
seems, therefore, a worthwhile goal.      

\break
\noindent{\bf 2 \hskip 0.2pc A Renormalizable Quantum Mechanical System} 
\vskip 1pc
It has been recognized for some time (see \[thorn],\[huang],
\[gupta],\[manuel]) that renormalizable ultraviolet 
divergences are not restricted to quantum field theories, but can occur as well
even in nonrelativistic quantum mechanics. A scale invariant Hamiltonian that
admits a negative energy bound state necessarily obtains a continuum of
negative energy states extending down to arbitrarily negative energies such
that there is no ground state. The system, without renormalization, is
thus unstable, and ill-defined.

The example we treat is representative of this situation. The attractive 
Dirac delta function potential in two dimensions with nonrelativistic kinetic
energy has been treated (see \[thorn],\[huang],\[manuel]) by the conventional 
renormalization recipe:
regularize with an ultraviolet cutoff, allow the coupling constant to run 
(depend on the cutoff), and remove the cutoff keeping some physical observable
such as the ground-state energy fixed. In this way divergences are removed, and
the dimensionless coupling constant characterizing the system is traded for a
dimensionful parameter, such as the ground state energy. This prescription may 
be administered either perturbatively (order by order in the coupling constant)
or nonperturbatively, but in either case the philosophy is the same: solve
a regularized, non-physical, system first, then take limits of the solutions to
obtain physical results.

Whether or not a direct renormalized description of the system to be solved
can be given to avoid such limiting procedures and non-physical intermediate
results is then a natural question. Here we give such a description
of the delta function potential in two dimensions, first in the Hamiltonian
and then in the path integral pictures. The result is a better understanding
of the role of the domain of the Hamiltonian in the former, and the description
of an interesting alternative to the Wiener measure in the latter. 
 
Our starting point {\it could be} the Hamiltonian:
\beq
        H_g = -\Delta - g\delta^2(\bar x)
\eeq where $\Delta$ is the two dimensional Laplacian, and $g$ is a positive
dimensionless number. In momentum space the Schrodinger equation is then:
\beq
p^2 \Psi(\bar p) - {g\over {(2\pi)^2}}\int d^2p \Psi(\bar p) =
\lambda \Psi(\bar p)                                      
\eeq\label{schro}\noindent Due to scale invariance, this equation admits bound
states for any energy,
$\lambda$, less than zero. They have the simple form which follows from a
rearrangement of \(schro):
\beq
\Psi(\bar p) = {{g\over{(2\pi)^2}}{\int d^2p \Psi(\bar p)}\over
{p^2 + |\lambda|}}
\eeq
However, integrating this equation over momentum space reveals a problem: $\int
d^2p \Psi(\bar p)$ is not finite. There is a logarithmic divergence in the
integral at high momenta. One way to cure this illness is to regularize by
introducing a large momentum cutoff $\Lambda$ and allowing the coupling
constant $g$ to depend on $\Lambda$ in the way which keeps the bound state
energy constant as $\Lambda$ is removed (taken to infinity). Integrating the
Schrodinger equation up to the cutoff $\Lambda$ gives:
\beq
1 = {g(\Lambda) \over {2\pi}} \int_0^{\Lambda} {pdp \over 
{p^2 + |\lambda_0|}}
\eeq which has the problem that the integral diverges logarithmically if
$\Lambda \to \infty$. The proper dependence of $g$ on $\Lambda$ to keep
$\lambda_0$ fixed is:
\beq
g(\Lambda) = {2\pi \over {\ln ({{\Lambda^2}\over {|\lambda_0|}} + 1)}}
\eeq\label{g}
The choice of $\lambda_0 (<0)$ is arbitrary, but must be made, and picks out 
just one of the uncountable number of possible bound states to survive the
renormalization procedure. With this choice, the cutoff can be removed, and the
parameter $g$ disappears
from the problem, replaced by the new parameter $\lambda_0$, the energy of the
single bound state of the system. All physical observables (e.g. scattering
amplitudes) may be calculated by solving this system with the cutoff in place,
and then taking the limit $\Lambda \to \infty$ with $g$ replaced by the 
expression in \(g). 

Nonperturbative and successful as this method is, we might ask for a direct
description in the Hamiltonian picture wherein the cutoff $\Lambda$ need not 
appear at all. If such a finite description of a renormalizable system is
possible, it should be easiest to find it here in our simple case.
This direct description may improve our understanding of the role of
renormalization and possibly serve as a guide to finding a similar point
of view in the more complicated case of quantum field theory.

The system we have described is asymptotically free. This means that the
coupling constant, $g$, goes to zero as $\Lambda$ is taken to infinity.
Nonetheless, if we take this limit in the momentum space Hamiltonian, with the
dependence of $g$ an $\Lambda$ given above, the Hamiltonian is not just $p^2$.
An interaction term survives, and the Schrodinger equation becomes:
\beq
p^2 \Psi(\bar p) - \lim_{p \rightarrow \infty} p^2 \Psi(\bar p) =
\lambda \Psi(\bar p)                                      
\eeq\label{renschro}

The renormalized Hamiltonian in \(renschro) appears to depend on no parameters.
However, for the Hamiltonian to be self-adjoint, its domain must be specified
carefully. The domain here is determined by the bound state energy (the
parameter of the theory) and consists of wavefunctions satisfying:
\beq
\int d^2p (\Psi(\bar p) - {{\eta_{\Psi}}\over {p^2 + \mu^2}})=0
\eeq where $\lambda_0 = -\mu^2$ and $\eta_{\Psi}\equiv\lim_{p \rightarrow 
\infty}p^2 \Psi(\bar p)$.

In real space, this equation picks out wavefunctions that diverge
logarithmically at the origin, but are still square-integrable.
This Hamiltonian operator is the momentum space equivalent of the self-adjoint
extension of the two-dimensional Laplacian, which is described in \[albeverio].
The parameter $\mu^2$ can be any positive real number, each value corresponding
to a different self-adjoint operator. It is related to the parameter $\alpha$
of \[albeverio] by:
\beq
\ln (2/\mu^2) = 2\gamma + 4\pi\alpha
\eeq 
\noindent where $\gamma$ is Euler's constant. The bound state wavefunction is:
\beq
\Psi_{\lambda_0}(\bar p) = {1 \over {p^2 + \mu^2}}
\eeq and the zero angular momentum scattering states with energy $\lambda$ are:
\beq
\Psi_{\lambda}(\bar p) = {1\over k}\ln({\mu^2 \over k^2}) \delta(p-k)
+ {2 \over {k^2 - p^2}}
\eeq

\noindent The non-zero angular momentum scattering states are simply the free 
ones: plane waves in real space, delta functions in momentum space.

Thus we conclude that this asymptotically free interaction in quantum
mechanics, corresponds in momentum space to the unusual Hamiltonian in
\(renschro) and in configuration space to the free Hamiltonian with a boundary
condition requiring angular momentum zero wavefunctions to diverge at the
origin. 

That an asymptotically free renormalizable interaction can be specified 
directly in terms of the domain of the Hamiltonian in configuration space has 
been discovered
previously, in the context of the large-N limit of the 1+1-dimensional massless
non-abelian Thirring (or Gross-Neveu) model, \[henderson]. 

\break
\vskip 1.5pc
\noindent {\bf 3 \hskip .2pc  The Feynman-Kac Formula and Path Integral Picture}
\vskip 1pc
\noindent Generally, quantum mechanical probability amplitudes and expectation
values may be calculated in Euclidean time using the Feynman-Kac formula:

\beq
<\bar x_1|e^{-tH}|\bar x_2>\quad = 
\int d\mu_0[\bar x]_{\bar x_1,\bar x_2,t}\quad
e^{-\int_0^t V(\bar x(s))ds}
\eeq
where $d\mu_0[\bar x]_{\bar x_1,\bar x_2,t} = \delta^2(\bar x(0)-\bar
x_1)\delta^2(\bar x(t)-\bar x_2)d\mu_0[\bar x]$ with $d\mu_0[\bar x]$ being the
Wiener measure (with the endpoints $\bar x(0)$ and $\bar x(t)$ left
unspecified) of classical diffusion theory. In Euclidean time, then, the 
stochastic
nature of quantum mechanics is indistinguishable from the classical randomness
of diffusion. A quantum mechanical free particle in Euclidean time has
correlation functions e.g. that are the same as those of a classical particle
executing Brownian motion with diffusion constant $D=\hbar^2/2m$. A quantum
particle in an external potential in Euclidean time, in its ground state say,
also behaves as a diffusing classical particle, but one with the Wiener measure
modified by the multiplicative factor $e^{-\int_0^t V(\bar x(s))ds}$.

The free heat kernel is, in two dimensions, the familiar probability density
for a particle to arrive at $\bar x_2$ at time $t$ having started at time $0$ at
point $\bar x_1$:

\beq
P^{(0)}_t(\bar x_2|\bar x_1)=h^{(0)}_t(\bar x_2,\bar x_1)=
<\bar x_2|e^{-tH_0}|\bar x_1> =
\int d\mu_0[\bar x]_{\bar x_1,\bar x_2,t} = {1 \over {4\pi t}}e^{-(\bar x_1 -
\bar x_2)^2 \over 4t}
\eeq\label{freeprob}

\noindent Equivalently, for the Wiener process, the probability density 
function of $\bar x(t_2)-\bar x(t_1)$ is:
\beq
P_{\bar x(t_2)-\bar x(t_1)}(\bar
x)={{1}\over{4\pi(t_2-t_1)}} e^{-x^2\over{4(t_2-t_1)}}
\eeq\label{excursion}

\noindent The "reproducibility property" of heat kernels:
\beq
\int d^2y \quad h_{t_1}^{(0)}(\bar x, \bar y)h_{t_2}^{(0)}(\bar y, \bar z) =
h_{t_1+t_2}^{(0)}(\bar x,\bar z)
\eeq ensures that the conditional probability in \(freeprob) is consistent.

\vskip 1.5pc
\noindent {\bf 4 \hskip .2pc  The Prokhorov Theorem}
\vskip 1pc
\noindent Implicit in the discussion above is the assumption that there exists
a probability measure (the Wiener measure) on the space of continuous curves in
${\bf R^2}$ which yields the probabilities in \(freeprob) and \(excursion). A
complete assignment of probabilities on this space of paths, however, requires
that {\it all} possible events (to be defined shortly) be given probabilities
consistent with the logic of probability and set theory. It is by no means
obvious that the probabilities in \(freeprob) can be generalized to such a true
probability measure. It is the purpose of this section to describe the
Prokhorov Theorem, which provides a simple test to which the probabilities in
\(freeprob) can be subjected in order to establish the existence of such a
measure. The Wiener process will be seen to pass this test, and in
the next section the test is used to verify that our renormalized quantum
mechanical system has a path integral description in terms of a new measure
which we believe to be distinct from the Wiener one in a way that systems
with nonsingular potentials cannot be.

The setting that we require is a {\it probability space}, which is a triple,
$(\Omega,{\cal B},P)$, where $\Omega$ is a set (the {\it sample space}), ${\cal
B}$ is a {\it Borel algebra} (an algebra of subsets of $\Omega$ closed under
countable unions and complementations) whose members are the possible {\it
events}, and $P:{\cal B}\to {\bf R}$ is a {\it probability measure}, meaning
that it must have the properties $P(\Omega)=1$ and countable additivity, i.e.
\beq
P\Bigl(\bigcup\limits_n A_n\Bigr) = \sum\limits_n P(A_n)
\eeq if $A_n \cap A_m = 0, \forall n\not= m$.

If instead of $\cal B$ we identify $\cal A$ as an algebra closed only under
finite unions, and $p:{\cal A}\to {\bf R}$ is additive only for finite unions,
then we say that $p$ is an {\it elementary probability measure}.

The probabilities given in \(freeprob) specify an elementary probability
measure to the set of events of the form:
\beq
E = \lbrace \bar x | \bar x \in \Omega , (\bar x(t_1),...,\bar x(t_n))\in
B^{2n} \rbrace
\eeq\label{event}

\noindent where $B^{2n}$ is a Borel subset of $\bf R^{2n}$ and $n$ is finite. 
$\cal A$ here is then the set of all such events. Their probabilities are:
\beq
P(E) = {\int d^2x_1 ... \int d^2x_n}_{B^{2n}} P_{\bar x(t_1)...
\bar x(t_n)}(\bar x_1,...,\bar x_n)
\eeq\label{prob} 

\noindent where
\beq
P_{\bar x(t_1)...\bar x(t_n)}=h_{t_1}^{(0)}(\bar 0,\bar x_1)h_{t_2-t_1}^{(0)}
(\bar x_1,\bar x_2)...h_{t_n-t_{n-1}}^{(0)}(\bar x_{n-1},\bar x_n)
\eeq\label{multiprob}

\noindent for $t_1<t_2<...<t_n$. Clearly an event $E$ has many equivalent 
descriptions.
One can always add more times to the list in \(event) without placing
restrictions on $\bar x$ at these new times without changing the event in any
way. The reproducibility property of the heat kernel ensures that each of these
equivalent descriptions will be assigned the same probability.

Thus with ${\cal A} = \lbrace E \rbrace$, we have defined an elementary
probability measure on $\Omega$. By the Kolmogorov theorem (see e.g.
\[prokhorov]) if $p$ is an elementary probability measure on ${\cal A}\subset
\Omega$ then there exists a unique extension to a full probability measure $P$
on $\cal B(\cal A)$, the Borel algebra generated by $\cal A$, if and only if for
any sequence of sets $A_n \in \cal A$ having $A_{n+1} \subset A_n$ and where
$\bigcap_n A_n = \emptyset$, the measure has the property
$\lim_{n \rightarrow \infty}p(A_n) = 0$. 

Fortunately, for our purposes, there  is an equivalent, but simpler, test when
$\Omega$ is the space of continuous paths. Prokhorov's theorem,
\[prokhorov] and \[ito], says
that in this case the unique extension to a full probability measure exists if
and only if the elementary probability measure assigns event probabilites such
that, for some $t_0>0$, there exist constants $a>0, b>1$, and $c>0$ such that:
\beq
E(|\bar x(t_2)-\bar x(t_1)|^a) \le c|t_2-t_1|^b,\quad \forall |t_2-t_1|\le t_0
\eeq\label{bound}

\noindent The Wiener process, by \(excursion), is scale-invariant, and therefore
satisfies this bound by taking $a=4$ and $b=2$.

We give an outline of the proof of Prokhorov's theorem in the Appendix. In the
next section we find the heat kernel of our renormalized quantum mechanical
system, and use this to define an elementary probability measure analagous to
the Wiener one. Then we show that this elementary measure also can be extended
to a full probability measure, thus constructing the path integral appropriate
for our system.

\vskip 1.5pc
\noindent {\bf 5 \hskip .2pc  The Renormalized Path Integral}
\vskip 1pc

As shown in Section 2, the renormalized delta function Hamiltonian is a
self-adjoint extension of the two-dimensional Laplacian. Eigenfunctions in
configuration space satisfy the free Schrodinger equation, but with a singular
boundary condition at the origin. The heat kernel, then, satisfies the heat
equation, with this same boundary condition at $x_i \to 0, i=1,2$:
\beq
-\Delta_{\bar x_i}h_t(\bar x_1,\bar x_2) = -{{\pdr h_t(\bar x_1,\bar
x_2)}\over {\pdr t}}
\eeq with
\beq
h_0(\bar x_1,\bar x_2)=\delta^2(\bar x_1 - \bar x_2), \quad h_t(\bar x_1,\bar
x_2)\sim C\ln \beta x_i, \quad x_i\to 0,\quad i=1,2
\eeq where $\ln ({\mu^2\over \beta^2})=2(\ln2 - \gamma)$. We can get the
explicit solution by first constructing the resolvent with the proper boundary
conditions:
\beq
G_{\lambda}(\bar x_1,\bar x_2) \equiv <\bar x_1|{1\over {-\Delta - k^2}}
|\bar x_2>
={1\over{2\pi}}K_0(k|\bar x_1 - \bar x_2|)+{1\over{2\pi}}
{{K_0(kx_1)K_0(kx_2)}\over {\ln ({k\over \mu})}}
\eeq\label{resolvent} 

\noindent where $\lambda = k^2$. The first term is the free resolvent,
$G_{\lambda}^{(0)}$. The second term is required to
achieve the boundary conditions for small $x_1$ and $x_2$. The resolvent is the
Laplace transform of the heat kernel, so we find that the heat kernel consists
of the free one plus a term which is a convolution of free heat kernels and
another function, $\nu$:
\beq
h_t(\bar x_1,\bar x_2) = h_t^{(0)}(\bar x_1,\bar x_2) + 4\pi\mu^2\int_0^t ds 
\int_0^s ds' h_{t-s}^{(0)}(\bar x_1,\bar
0)\nu(\mu^2(s-s'),-1)h_{s'}^{(0)}(\bar 0,\bar x_2)
\eeq\label{renkernel}

\noindent with $\nu$ being the function
\beq
\nu(t,a)\equiv \int_{a}^\infty ds {{t^s}\over{\Gamma(s+1)}}
\eeq

We would like to use this heat kernel to define, as in \(multiprob), an
elementary probability measure on the sets in \(event), and then extend this 
using the Prokhorov
theorem to a full probability measure, thus yielding a path integral
description of our renormalized Hamiltonian. The first problem we encounter is 
that, as when the free Hamiltonian, $H_0$, is modified by
the addition of a potential, no longer is $\int d^2x_2
h_t(\bar x_2,\bar x_1) = 1$, so we cannot interpret $h_t(\bar x_2,\bar x_1)$ as
a probability density (i.e. as $P_t(\bar x_2|\bar x_1)$). A normalization is
necessary. As in the case $V\ne0$ we can utilize the positive, normalizable,
ground state $\Psi_{\lambda_0}(\bar x)$ to define:
\beq
P_t(\bar x_2|\bar x_1)=e^{-\mu^2 t}{{\Psi_{\lambda_0}(\bar
x_2)}\over{\Psi_{\lambda_0}(\bar x_1)}} h_t(\bar x_2,\bar x_1)
\eeq which is a normalized probability density, still has the reproducibility
property, and for small times is $\sim h_t(\bar x_2,\bar x_1)$. In our case,
\beq
\Psi_{\lambda_0}(\bar x)= {{\mu}\over{\sqrt{\pi}}}K_0(\mu x)
\eeq

The elementary probability measure we want is then given by \(prob) and
\(multiprob), with $h_t^{(0)}$ replaced by $P_t$. The diffusion generated by 
this probability gives correlation functions and
expectation values which are the quantum mechanical ground state correlation
functions and expectation values in Euclidean time. 

Now $P_t$, unlike $h_t^{(0)}$, is not scale-invariant, since the ground state
energy $-\mu^2$ sets a scale. It takes, therefore, some work to show that this
elementary probability measure satisfies the Prokhorov bound, and therefore has
a unique extension to a probability measure on the space of continuous
two-dimensional paths. The expectation value that we must bound can be written
as a sum of two terms:
\beq
E(|\bar x(t_2) - \bar x(t_1)|^a) =
E^{(0)}(|\bar x(t_2) - \bar x(t_1)|^a)+
E^{(1)}(|\bar x(t_2) - \bar x(t_1)|^a)
\eeq where the first term corresponds to the free part of $P_t$, and the second
term comes from the interaction part of $P_t$. We can bound these two postive
terms separately:
\beq
E^{(0)}(|\bar x(t_2) - \bar x(t_1)|^a)={\mu^2 \over \pi}e^{-\mu^2 t}
\int d^2x_1 d^2x_2 |\bar x_2 - \bar x_1|^a K_0(\mu x_1)K_0(\mu x_2)
h_t^{(0)}(\bar x_1,\bar x_2)
\eeq
Useful here is a power expansion of $K_0(\mu x_2)$:
\beq
K_0(\mu x_2) = K_0(\mu x_1) + R(x_1,x_2)
\eeq where for $|x_2-x_1|<\delta$, $\delta$ being small,
\beq 
|R(x_1,x_2)| = |-\mu K_1(\mu x_1)(x_2-x_1) + \dots| < A(\delta)
K_1(\mu x_1)|x_2-x_1|
\eeq 
$A(\delta)$ being positive and constant with respect to $x_1$ and $x_2$.
Inserting this bound on $K_0(\mu x_2)$ (note that $K_0$ and $K_1$ are positive
for positive arguments) yields the bound:
\beq
E^{(0)}(|\bar x(t_2) - \bar x(t_1)|^a) < C_1 e^{-\mu^2 t} (\mu^2 t)^{a/2}
            +C_2 e^{-\mu^2 t} (\mu^2 t)^{{a+1}\over 2}
\eeq where $C_1$ and $C_2$ are dimensionless constants. The other term in the
expectation value we need is
\beq
E^{(1)}(|\bar x(t_2) - \bar x(t_1)|^a)={\mu^2 \over \pi}e^{-\mu^2 t}
\int d^2x_1 d^2x_2 |\bar x_2 - \bar x_1|^a K_0(\mu x_1)K_0(\mu x_2)
h_t^{(1)}(\bar x_1,\bar x_2)
\eeq\label{e1}

\noindent defining $h_t^{(1)}$ to be $h_t - h_t^{(0)}$, the interaction term in
the heat kernel. In this expression we may use:
\beq
K_0(\mu x_1)K_0(\mu x_2) < C(\epsilon) (x_1x_2)^{-\epsilon}
\eeq 
where $\epsilon$ is any positive real number. Inserting this into
the expression for $E^{(1)}$ and taking the Laplace transform with respect to 
$t\equiv t_2 - t_1$ makes, using the resolvent formula, the integral on the 
right-hand side of \(e1)
\beq
{C(\epsilon) \over {2\pi \ln{k\over \mu}}} \int d^2x_1 d^2x_2
(x_1x_2)^{-\epsilon} |\bar x_2 - \bar x_1|^a K_0(kx_1)K_0(kx_2)
\eeq Conveniently, we can scale $\bar x_1$ and $\bar x_2$ to bring all the $k$
dependence outside the integral, giving us
\beq
{C(\epsilon,a)\over {k^{4+a-\epsilon}\ln({k\over \mu})}}
\eeq which has the inverse Laplace transform
\beq
{{2 C(\epsilon,a)}\over {\mu^{2+a-\epsilon}}} \nu(\mu^2 t,
{{a+2-\epsilon}\over 2})
\eeq

\noindent The function $\nu$ has an asymptotic expansion, \[bateman],:
\beq
\nu (\mu^2 t,p) = {(\mu^2 t)^p \over {\ln({1\over {\mu^2 t}})}}
(C_p + O(|\ln({1\over{\mu^2 t}})|^{-1}))
\eeq 
Using this provides a bound on $E^{(1)}$ for small $t$:
\beq
E^{(1)}(|\bar x(t_2) - \bar x(t_1)|^a)< C(\epsilon,a,t_0)
{{(\mu^2 t)^{{2+a-\epsilon}\over 2}}\over
{\ln({1\over{\mu^2 t}})}}, \quad \forall t \le t_0
\eeq
This result and the above bound on $E^{(0)}$ combine to prove that, as for the
Wiener measure, the Prokhorov bound is satisfied with the values $a=4$ and
$b=2$. This means that there is a unique probability measure, and hence path
integral, to describe our renormalized quantum mechanical system.
The Wiener measure in path integrals is then replaced by $d\mu_{\lambda_0}[\bar
x]$ where
\beq
h_t(\bar x_1,\bar x_2) = \int d\mu_{\lambda_0}[\bar x]_{\bar x_1,\bar x_2,t}
\eeq and the addition of a potential modifies this to:
\beq
h_t(\bar x_1,\bar x_2) = \int d\mu_{\lambda_0}[\bar x]_{\bar x_1,\bar
x_2,t}\quad e^{-\int_0^t V(\bar x(s))ds}
\eeq

Like the Wiener measure, $d\mu_{\lambda_0}[\bar x]$ is thus a measure on the
space of continuous paths in $R^2$. An ordinary, nonsingular, interaction
modifies the Wiener measure by the multiplicative factor 
$e^{-\int_0^t V(\bar x(s))ds}$, a functional which for small times approaches
unity, with corrections of $O(t)$. Under the influence of such interactions
particles, for small enough times, behave as free ones. In contrast, the 
measure $d\mu_{\lambda_0}[\bar x]$, corresponding to the renormalized delta 
function interaction, also approaches the Wiener measure as $t \to 0$, but the
corrections die much more slowly, being logarithmic rather than power law. This
interaction, though subtle in the way it breaks scale invariance, modifies
particle motion for very small times in a more profound way than does an
ordinary potential. We conjecture, but have not proved, that the mathematical
consequence of this is that, unlike the Wiener measure multiplied by the
functional $e^{-\int_0^t V(\bar x(s))ds}$, the measure 
$d\mu_{\lambda_0}[\bar x]$ not {\it absolutely continuous}, see e.g.
\[freedman], with respect to the free Wiener measure for arbitrarily small 
times.   
                   
It is also interesting to note the meaning (in the diffusion picture) of the
second term in \(renkernel). This term, which is positive, corresponds to the
probability that, rather than undergoing ordinary Brownian motion, the
particle in going from $\bar x_1$ to $\bar x_2$, first diffuses into the
neighborhood of the origin where it spends some random amount of time (the
distribution of this random delay being proportional to the function $\nu$)
before diffusing out again to its destination $\bar x_2$. In fact, keeping this
picture in mind provides an alternative way to arrive at the heat kernel,
$h_t$. Adding the probability of such an excursion to the origin amounts to
adding a term $f(\lambda)G_{\lambda}^{(0)}(\bar x,\bar 0)
G_{\lambda}^{(0)}(\bar 0,\bar y)$ to the free resolvent. $f(\lambda)$ is then
the Laplace transform of the random time delay distribution. The resulting
ansatz for the resolvent,
\beq
G_{\lambda}(\bar x,\bar y) = G_{\lambda}^{(0)}(\bar x,\bar y) +
f(\lambda)G_{\lambda}^{(0)}(\bar x,\bar 0)G_{\lambda}^{(0)}(\bar 0,\bar y)
\eeq
should then be required to have the reproducibility property, required for it
to be the resolvent of {\it some} operator:
\beq
\int d^2y G_{\lambda}(\bar x,\bar y)G_{\lambda}(\bar y,\bar z) =
{-d\over {d\lambda}}G_{\lambda}(\bar x,\bar z).
\eeq
Making this requirement, and using 
$G_{\lambda}^{(0)}(\bar x,\bar y)={1\over {2\pi}}K_0(k|\bar x - \bar y|)$,
yields a first-order nonlinear differential equation for $f$:
\beq
f'(\lambda) = {{-f^2(\lambda)}\over {4\pi \lambda}}
\eeq
which can be integrated to give $f(\lambda)=2\pi/\ln({\sqrt{\lambda}\over
\mu})$, where here $\mu$ appears as an integration constant. This result 
matches the expression given for the resolvent in \(resolvent).

Consideration then of
nonrelativistic quantum particles acting through the renormalized delta
function potential is equivalent to the treatment of classical particles
undergoing a modified Brownian diffusion with enhanced probability for the
particles to stick together for some time before going their separate ways. The
possibility of such a "sticky diffusion" in two dimensions may be of interest
even in classical diffusion theory or in condensed matter physics.

\vskip 1.5pc
\noindent {\bf \hskip .2pc  Appendix}
\vskip 1pc

\noindent Here we provide a roadmap to the proof of the Prokhorov theorem as
given in \[ito]. The idea is to show that the "Prokhorov bound", \(bound),
implies the conditions of the Kolmogorov theorem. That is, if the elementary
probability measure implies \(bound),
and we consider a set of events $\lbrace A_n \rbrace$ of the form \(event) such
that $A_{n+1}\subset A_n$ and $p(A_n)>\epsilon>0$ then $\bigcap_n A_n \not=
\emptyset$.

Now, by adding times at which the particle position is unrestricted, we can
always put such a sequence of events into a standard form:
\beq
A_n = \lbrace \bar x| (\bar x(t_1^{(n)}),...,\bar x(t_{n2^{n+1}}^{(n)}))\in B^n
\rbrace
\eeq\label{A_n}\noindent Since $p$ is a probability measure when restricted to
sets dependent on a fixed set of times, we can assume that each $B^n$ is a
compact set.
Also, by choosing $n$ large enough we can make $t_i^{(n)}-t_{i-1}^{(n)} < 2^{-n}
< t_0$. The Prokhorov bound and Chebyshev's inequality then give, for any
$\delta >0$
\beq
p(\bar x||\bar x(t_i^{(n)})-\bar x(t_{i-1}^{(n)})|^a \ge 
|t_i^{(n)}-t_{i-1}^{(n)}|^{\delta a}) \le c|t_i^{(n)}-t_{i-1}^{(n)}|^{b-\delta
a}
\eeq Let $\lambda \equiv b - \delta a - 1$. Using this, $p(A_l)>\eps$, and 
DeMorgan's law we can show that $p(E_l)>{\epsilon \over 2}$ where $\lbrace E_l
\rbrace$
are the events
\beq
E_l=A_l\bigcap\Biggl(\bigcap\limits_{n=m_0}^l\bigcap\limits_{i=2}^{n2^{n+1}}
\lbrace \bar x||\bar x(t_i^{(n)})-\bar x(t_{i-1}^{(n)})| < 
|t_i^{(n)}-t_{i-1}^{(n)}|^{\delta } \rbrace\Biggr)
\eeq where $m_0$ is taken large enough that $2c\sum_{n=m_0}^\infty
n2^{-n\lambda}<{\epsilon \over 2}$. So these sets are non-empty. The event 
$E_l$ includes all paths which
belong to $A_l$ and in addition have the property that in each time division
between $m_0$ and $l$, i.e. for $m_0<n<l$, the distance travelled is bounded
above by $\Delta t^\delta$.

Now it is clear that $E_{l+1}\subset E_l\subset A_l$ so that showing
$\bigcap_{l=m_0}^\infty E_l \ne \emptyset$ is sufficient to prove 
$\bigcap_{l=m_0}^\infty A_l \ne \emptyset$. To show the former, first pick from
each set $E_l$ a path, $\bar x_l(t)$, which is linear in each time segment,
$[t_{i-1}^{(l)},t_i^{(l)}]$. Geometry and the definition of $E_l$ imply that
for $t_i^{(l)}\le t \le s \le t_j^{(l)}$
\beq
|\bar x_l(t)-\bar x_l(s)| \le 
K|t_i^{(l)}-t_{j}^{(l)}|^{\delta }
\eeq\label{xbound}

\noindent for a constant $K>0$. Now the paths $\bar x_{l+p}$ belong to $A_l$ 
for all $p>0$, and therefore $(\bar x_{l+p}(t_1^{(l)}),...,\bar
x_{l+p}(t_{l2^{l+1}}^{(l)})) \in B^l$. $B^l$ is compact. Therefore this
sequence, indexed by $p$, has a limit point in $B^l$. This is true for all
$l$. Thus we can by the diagonalization method extract a subsequence, $\lbrace 
\bar y_n \rbrace$ such that
$\bar y_n(t_i^{(l)})$ converges as $n\to \infty$ for all $i$ and $l$. Now, if
$\eta$ and $\tau$ are given, we can choose $n_0$ large enough such that
$t_i^{(n_0)} \le \tau \le t_{i+1}^{(n_0)}$ and $|t_i^{(n_0)}-t_{i+1}^{(n_0)}|
< 2^{-n_0} < {\eta \over 2}$. Then choosing $l$ and $m$ large enough, we can
show by triangle inequalities that $|\bar y_l(\tau)-\bar y_m(\tau)|<A{\eta \over
2}$ for some positive constant $A$. This is true for any $\tau \in
[t_i^{(n_0)},t_{i+1}^{(n_0)}]$. Thus the limiting function, say $\bar y(t)$, 
exists $\forall t\in {\bf R}$, and \(xbound) ensures that $\bar y(t)$ is
continuous. $\bar y(t)$ has the property $(\bar y(t_1^{(l)}),...\bar
y(t_{l2^{l+1}}^{(l)})) \in B^l, \forall l$, so that $\bar y \in \bigcap_{l\ge
m_0}E_l \ne \emptyset$, completing the proof.

\vskip 1.5pc
\noindent {\bf References}\hfill
\vskip 1pc 
\def\ni{\noindent}

\ni\thorn. C. Thorn, Phys. Rev. D19, 639 (1979).
 
\ni\huang. K. Huang, "Quarks, Leptons and Gauge Fields", World Scientific,
Singapore (1982).
 
\ni\gupta. K.S. Gupta and S.G. Rajeev, Phys. Rev. D48, 5940 (1993).

\ni\manuel. C. Manuel and R. Tarrach, Phys. Lett. B328, 113 (1994).

\ni\albeverio. S. Albeverio, F. Gesztesy, R. Hoegh-Krohn, and H. Holden,
"Solvable Models in Quantum Mechanics", Springer-Verlag, New York, NY (1988)

\ni\henderson. R.J. Henderson and S.G. Rajeev, Intl. J. Mod. Phys. A 10, 3765
(1995).

\ni\prokhorov. Y.V. Prokhorov, Theory of Prob. and its Applic. 1, 157 (1956).
 
\ni\ito. "Lectures on Stochastic Processes", by K. Ito, notes by M.Rao, TIFR
(1961)

\ni\bateman. Erdelyi, A. (Editor), The Bateman Manuscript Project, Higher 
Transcendental Functions, Vol III, p217, McGraw-Hill, New York, NY (1955)

\ni\freedman. D. Freedman, "Brownian Motion and Diffusion", Holden-Day, Inc.,
San Francisco, CA (1971) 

\bye